\documentclass[12pt]{article}
\usepackage{latexsym}
\usepackage{amssymb, amsmath, xspace, lscape,  latexsym}

\renewcommand{\dim}{\mathrm{dim \,}}

\renewcommand{\le}{\leqslant}

\newcommand{\pr}{\prime}
\newcommand{\bea}{\begin{eqnarray}}
\newcommand{\eea}{\end{eqnarray}}
\def\beq#1#2\eeq{
        \begin{equation}
        \label{#1}
            #2
        \end{equation}}

\newcommand{\mref}[1]{(\ref{#1})}

\newcommand{\px}{\partial_x}
\newcommand{\py}{\partial_y}
\newcommand{\pt}{\partial_t}

\newcommand{\al}{\alpha}
\newcommand{\bt}{\beta}
\newcommand{\ga}{\gamma}

\newcommand{\Z}{\mathbb Z}

\renewcommand{\tilde}{\widetilde}
\newcommand{\Int}{\int_{-\infty}^{\infty}}

\def\btheor#1\etheor{
        \begin{theor}
            #1
        \end{theor}
    }

    \def\bsled#1\esled{
        \begin{sled}
            #1
        \end{sled}   }

\newtheorem{theorem}{Theorem}

\newtheorem{prop}{Proposition}

\newcommand{\nad}[2]{\genfrac{}{}{0pt}{}{#1}{#2}}

\def\hm#1{#1\nobreak\discretionary{}{\hbox{\m@th$#1$}}{}}
\def\mi#1{\discretionary{\hbox{\m@th$#1$}}{\hbox{\m@th$#1$}}{}}

\vspace{4ex}
\begin{document}
\title{\bf Painlev\'e IV and degenerate Gaussian Unitary Ensembles}
\author{Yang Chen$^{\dag}$\\
        Department of Mathematics\\
        Imperial College London\\
        180 Queen's Gates\\
        London SW7 2BZ UK\\
        \\
        \\
        M. V. Feigin$^{*}$\\
        Department of Mathematics\\
        University Glasgow\\
        University Gardens\\
        Glasgow G12 8QW UK }
\date{28-06-2006}
\maketitle
\begin{abstract}
We consider those Gaussian Unitary Ensembles where the eigenvalues
have prescribed multiplicities, and obtain joint probability density
for the eigenvalues. In the simplest case where there is only one
multiple eigenvalue $t$, this leads to orthogonal polynomials with
the Hermite weight perturbed by a factor that has a multiple zero at
$t.$ We show through a pair of ladder operators, that the diagonal
recurrence coefficients satisfy a particular Painlev\'{e} IV
equation for any real multiplicity. If the multiplicity is even they
are expressed in terms of the generalized Hermite polynomials, with
$t$ as the independent variable.
\end{abstract}
\noindent
$^{\dag}$ychen@ic.ac.uk

\noindent
$^{*}$m.feigin@maths.gla.ac.uk
\vfill\eject

\setcounter{equation}{0}
\section{Introduction}

 Random matrix ensembles originally conceived to explain the statistical properties of the
energy levels in heavy nuclei \cite{W} has recently seen
applications in transport in disordered systems, string theory and
various areas of pure and applied mathematics. In addition to
classical quantities of interest such as the correlation functions,
the average of the product of characteristic polynomials of random
matrices were under investigation starting from Brezin-Hikami paper
\cite{BH} (see also \cite{BS} and references therein).

From the Painlev\'e equations point of view the average of a power
of characteristic polynomial in Gaussian unitary ensemble gives a
$\tau$-function of the rational solution of Painleve IV equation.
For the integer powers this can be seen from the original
Kajiwara-Ohta determinant formula for the rational solutions of PIV
(\cite{KO}, c.f. \cite{NY}) and it was later explored by Forrester
and Witte \cite{FW}.

In this paper we consider the degenerate gaussian unitary ensembles.
That is we restrict ourselves to the nonlinear subspace of Hermitian
matrices having prescribed spectrum degeneracy. Various statistical
properties on eigenvalues of such  matrices can be asked. The first
natural question we answer is the determination of the joint
probability density of the eigenvalues when these have some
multiplicity.
It happens that as in the case of classical ensembles (see
\cite{MM}) the joint probability density has the form of product of
pairwise differences between the different eigenvalues taken in the
powers depending on the multiplicities. Thus we naturally arrive to
considering orthogonal polynomials with Hermite weight perturbed
(multiplied) by a product of linear factors.

These types of weights also appear in the random matrix theory in
consideration of the averages of characteristic polynomials (see
\cite{BH}, \cite{BDS}), although in that case the zeroes of these
factors are the external variables to the matrices of the ensembles.
We note that an orthogonal circular random matrix ensemble with
fixed degenerate eigenvalue at 1 was considered by Snaith in
\cite{S} in conjectural relation to number theoretic questions on
$L$-functions of elliptic curves. More general Jacobi circular
ensembles were studied recently in \cite{D}. A general approach to
the joint probability density of ensembles of various type was
suggested recently in \cite{AWY}.

We are also motivated by the theory of Calogero-Moser-Sutherland
systems. The ground states of these systems at appropriate
interaction parameter coincide with the joint probability densities
of eigenvalues in the classical ensembles. The joint probability
density for degenerate ensembles coincides with the factorized wave
function for the appropriate multi-species generalisation of
Calogero--Moser problem considered in \cite{F}. This type of
generalisation is integrable in the case of two types of particles
\cite{CFV}, \cite{SV1}. More remarkably, Sergeev and Veselov showed
that the corresponding quantum Hamiltonian can be obtained by
applying a restriction procedure on the Calogero-Moser-Sutherland
Hamiltonian in the infinite dimensional space to the appropriate
discriminant \cite{SV2}. We plan to elaborate these relations in
future.

In the context of orthogonal polynomials, perturbations of the
standard weights such as the Jacobi weight by special factors is an
important topic of investigation, where the problem is the
determination of the recurrence coefficients from the weights (see
\cite{Mag3}, \cite{Mag2} and the references therein). In particular,
it was noted by Magnus in \cite{Mag1} that often the variations lead
to the recurrence coefficients which are solutions to the nonlinear
equations. In some cases the appearance of Painlev\'e IV for the
certain exponential weights was established \cite{Mag1}. More
recently, it was shown in \cite{chen3} that the diagonal recurrence
coefficient associated with the Hermite weight perturbed by special
discontinuous factor satisfies a particular Painlev\'e IV.

In this paper we show that when the Hermite weight is perturbed by a
linear factor having multiple zero the diagonal recurrence
coefficients satisfy a particular two parameter Painlev\'e IV
equation. This property in fact holds for an arbitrary real power of
the linear factor. Our approach is direct, it is based on an
extension of the ladder operators technique developed in
\cite{chen1}. In Section 3 we describe this method, suitable for
orthogonal polynomials where the weight has isolated zeros, in
particular we derive a pair of fundamental compatibility conditions
$(S_1)$ and $(S_2)$. In Section 4 we make use of these to generate
non-linear difference equations satisfied by the recurrence
coefficients. These difference equations when combined with the Toda
equations give a PIV equation  satisfied by the recurrence
coefficients $\alpha_n$.

In the cases of the weights arising from degenerate gaussian
ensembles having one multiple eigenvalue $t$ of degeneracy $K$ the
multiplicity of the linear factor in the weight is $2K$. In this
case the recurrence coefficients are rational as functions of $t$.
The theory of rational solutions to PIV \cite{NY} results in the
expression of the recurrence coefficients through the generalized
Hermite polynomials.

We also mention here that the Hankel determinants associated to the
Hermite weights perturbed by a factor are related to the
Hankel determinants of the Hermite weight with the addition of
$\delta$-function and its derivatives \cite{gru}. For an alternative
derivation using Heine's multiple integral see \cite{chen4}.

\setcounter{equation}{0}
\section{Non-generic random matrices}

Let ${\cal H}_N$ be the space of Hermitian matrices of size $N$ and
let $m=(m_1, m_2, \ldots, m_k)$ be a partition of $N$. Consider the
(nonlinear) subspace ${\cal H}^m_N$ in ${\cal H}_N$ consisting of
matrices having the eigenvalues with prescribed multiplicities
$m_1,\ldots,m_k$. That is we suppose the spectrum $\{\lambda_1,
\ldots,\lambda_N\}$ of an arbitrary element $A\in {\cal H}_N^m$ has
the multiplicities described below: \bea
\mu_1&=&\lambda_1=\ldots=\lambda_{m_1},\nonumber\\
\mu_2&=&\lambda_{m_1+1}=\ldots=\lambda_{m_1+m_2}, \nonumber\\
&.&\nonumber\\
&.&\nonumber\\
\mu_k&=&\lambda_{m_1+m_2+\ldots+m_{k-1}+1}=\ldots=\lambda_N,
\eea
where we have renamed the eigenvalues as $\mu_1,\ldots,\mu_k$ without repetitions.

As every Hermitian matrix $A$ is diagonalizable, we have
\beq{spc}
A=U \Lambda U^{-1},
\eeq
 where
$\Lambda=\textrm{diag}(\lambda_1, \ldots, \lambda_N)$, and $U$ is
unitary. The matrix $U$ is constructed out of a certain orthonormal
basis where $A$ becomes diagonal. Such a basis is defined up to
unitary transformations leaving the eigenspaces invariant. Therefore
$U$ is determined as an element of the homogenous space \beq{coset}
U \in U(N)/U(m_1)\times\ldots\times U(m_k), \eeq where the direct
product $U(m_1)\times\ldots\times U(m_k)$ of unitary matrices of
orders $m_1,\ldots, m_k$ is embedded into $U(N)$ as diagonal block.
More precisely, in order to determine $U$ uniquely, we also assume
that the eigenvalues $\mu_i, \mu_j$ having equal multiplicities
$m_i=m_j$ are such that $\mu_i < \mu_j$ if $i<j$.

 Although the subspace ${\cal H}_N^m,$ of ${\cal H}_N$ is a measure zero set,
we may nonetheless construct a natural probability measure of
the matrices lying in it. The metric
\beq{tr}
(ds)^2 = \textrm{tr} (dH^{*} dH)
\eeq
is well-defined in the subspace ${\cal H}_N^m$. Therefore this metric
also naturally defines a measure on the subspace ${\cal H}_N^m$, via
the Riemann volume formula. It happens, just
like in the case of Hermitian matrices with distinct eigenvalues,
with the spectral decomposition \mref{spc},
the measure on ${\cal H}_N^m$ is a product of
a measure on the eigenvalues and a measure on the homogeneous space
\mref{coset}.

\begin{prop}\label{jd}

The metric \mref{tr} restricted to the subspace ${\cal H}_N^m$ has
the form
$$
(ds)^2=\sum_{i=1}^k m_i d \mu_i^2 + 2 \sum_{1\le i < j \le k}
(\mu_i-\mu_j)^2 (ds_{ij})^2,
$$
where
\beq{sij}
 (ds_{ij})^2= \sum_{\nad{m_1+\ldots+m_{i-1}+1 \le \al
\le m_1+\ldots+m_i}{m_1+\ldots+m_{j-1}+1 \le \bt \le
m_1+\ldots+m_j}}(U^{-1}dU)_{\al \bt}(\overline U^{-1}\overline{d
U})_{\al\bt}. \eeq
 The corresponding volume form on ${\cal H}_N^m$ is
\beq{mes}
 d\mu = \prod_{1\le i < j \le k}
(\mu_i-\mu_j)^{2 m_i m_j}\;\prod_{i=1}^k d \mu_i\;d \nu(U),
 \eeq
where $d\nu(U)$ is invariant measure on the homogeneous space
\mref{coset}.
\end{prop}
{\bf Proof.} From decomposition \mref{spc} we obtain
$$
dA = U \left( d \Lambda + U^* dU \Lambda - \Lambda U^* dU\right)
U^*.
$$
Then the metric \mref{tr} can be rewritten as follows:
$$
(ds)^2=\textrm{tr}\left( (d \Lambda)^2 + 2(\delta U \Lambda - \Lambda
\delta U)d\Lambda +(\Lambda \delta U)^2+(\delta U \Lambda)^2
-2\delta U \Lambda^2 \delta U  \right),
$$
where $\delta U:= U^{-1} dU$ and we have used the cyclic property of
the trace. Simplifying this further we arrive at
$$
(ds)^2=\textrm{tr}\left((d\Lambda)^2+ 2\sum_{i\ne j} \left(\lambda_i
\lambda_j\delta U_{ij} \delta U_{ji}
 - \lambda^2_i\delta U_{ij} \delta
U_{ji}\right) \right),
$$
and since $\delta U$ is anti-Hermitian we get (c.f., e.g.,
\cite{FGZ}, \cite{HW}) that the above reduces to
$$
(ds)^2 = \sum_{i=1}^N (d \lambda_i)^2 + 2 \sum_{1\le i<j\le N}
(\lambda_i-\lambda_j)^2 \delta U_{ij} \overline{\delta U_{ij}}.
$$
Recalling the degeneracy conditions (2.1) we note that some of the
terms vanish and the restricted metric takes the form \beq{met}
 (ds)^2=\sum_{i=1}^k m_i(d\mu_i)^2
+ 2 \sum_{1\le i < j \le k} (\mu_i-\mu_j)^2(ds_{ij})^2,
\eeq
 where
$ds_{ij}$ is defined in \mref{sij}. The second sum in \mref{met} is
well defined in the homogeneous space.

 To determine the corresponding measure we fix locally the section of the representatives
of the coset classes and consider coordinates $u_{\al\bt}$ such that
$du_{\al\bt}= (U^{-1} d U)_{\al\bt}$, where the indices $\al<\bt$
are such that $(\al\bt)\notin \Delta.$ Here $\Delta$ is the diagonal
block containing $U(m_1)\times\ldots\times U(m_k)$. Such local
coordinates $u_{\al\beta}$ will exist if the section is chosen to
satisfy $\delta U_{ij}=0$ when $(ij)\in\Delta$. Then taking the real
and imaginary parts $\Re u_{\al \bt}, \Im u_{\al \bt},$ as real
coordinates the metric \mref{met} becomes a diagonal metric $g_{ii}$
and the term $2(\mu_i-\mu_j)^2$ appears $2 m_im_j$ times along the
diagonal. From the Riemann volume formula, the measure corresponding
to \mref{met} is
$$
\prod_{i=1}^M \sqrt{|g_{ii}|}\prod_{i=1}^{k} d\mu_i
\prod_{\nad{(\al\bt)\notin\Delta}{\al<\bt}} d\Re u_{\al\bt} d\Im
u_{\al\bt},
$$
where
$$
\prod_{i=1}^{M}{\sqrt {|g_{ii}|}}=\prod_{1\leq i<j\leq k}2^{m_i m_j}
(\mu_i-\mu_j)^{2m_im_j} \prod_{i=1}^k m_i^{1/2},
$$
and $M=\dim {\cal H}^m_N$. Thus we obtain the result \mref{mes} with
the measure $d\nu(U)$ given by
$$
 d\nu(U)=\prod_{i=1}^k m_i^{1/2}
\prod_{\nad{(\al\bt)\notin\Delta}{\al<\bt}}
 (U^{-1}d U)_{\al\bt}(\overline{U}^{-1}\overline{d U})_{\al\bt}.
$$

{\bf Remark 1.} A large class of generalized random matrix ensembles
was recently considered in \cite{AWY} where a formula for joint
probability density of the eigenvalues was obtained. Expression
\mref{mes} may be obtained from that work.

{\bf Remark 2.} One way to generalize Proposition 1 is to consider
real symmetric matrices with multiple spectrum, then angular
variables are given by a factor in the orthogonal group. Same
arguments as above lead to the following joint probability density
of eigenvalues
$$
 \prod_{1\le i < j \le k} (\mu_i-\mu_j)^{m_i
m_j} \prod_{i=1}^k d \mu_i.$$ Another possibility is to consider
degenerate circular ensembles, that is unitary (or other) ensembles
with given spectrum multiplicities. In this case the calculation of
joint probability density results in taking the appropriate powers
of nontrivial Cartan roots.

It is a well-known result of Random Matrix theory \cite{MM} that the
partition function of any unitary invariant matrix ensemble defined
by the multiple integral,
\beq{parti}
\Delta_N[w]:=\frac{1}{N!}\int_{a}^{b}...\int_{a}^{b} \prod_{1\leq
i<j\leq N}(x_i-x_j)^2\prod_{k=1}^{N}w(x_k)dx_k, \eeq has the
alternative representations, namely \bea
\Delta_N[w]&=&\det\left(\int_{a}^{b}x^{i+j}w(x)dx\right)_{i,j=0}^{N-1}\\
&=&\det\left(\int_{a}^{b}p_{i}(x)p_{j}(x)w(x)dx\right)_{i,j=0}^{N-1},
\eea where $p_l(x)$ is an arbitrary monic polynomials of exact
degree $l.$ Now if we orthogonalise these with respect to the weight
$w$ over $[a,b],$ namely,
$$
\int_{a}^{b}p_i(x)p_j(x)w(x)dx=h_i\delta_{i,j},
$$
where $h_i,\;i\in\mathbb{N}$ is the square of the $L^2$ norm, then
\mref{parti} becomes, \beq{prod} \Delta_N[w]=\prod_{j=0}^{N-1}h_j.
\eeq For the generic Gaussian Unitary Ensembles, $w(x)=\exp(-x^2),$
$x\in\mathbb{R}.$ In the case of a single degenerate eigenvalue $t$
with $K$ fold degeneracy and the rest, $n$ eigenvalues are distinct,
such that $N=n+K$, we find, by relabeling,
$\mu_1=t,\;\mu_2=x_1,\ldots,\mu_k=x_n,$ the partition function
reads, \beq{det} \Delta_{n+K} =\Int \textrm{e}^{-Kt^2}D_n(t)dt, \eeq
where \beq{sdet} D_n(t)=\frac{1}{n!}\Int..\Int\prod_{1\leq i<j\leq
n}(x_i-x_j)^2 \prod_{l=1}^{n}(x_l-t)^{2K}\textrm{e}^{-x_l^2}dx_l.
\eeq We note the partition function expressions \mref{det},
\mref{parti} are defined here up to constant multiples that come
from the integration over the corresponding homogeneous spaces.

The weight of orthogonal polynomials associated with integral
\mref{sdet} is the Hermite weight multiplied by an isolated zero,
that is, $$ w(x;t)=\exp(-x^2)|x-t|^{2K},\;\;x,t\in\mathbb{R}.
$$

Other crucial characteristics of Random Matrix ensembles are the
correlation functions of the eigenvalues. These are obtained by
calculating the partition function type integrals \mref{parti} when
some of the eigenvalues are fixed. In the case of single degenerate
eigenvalue those correlation functions that involve the multiple
eigenvalue coincide with the averages of the powers of
characteristic polynomial for the appropriate standard Gaussian
unitary ensemble, as it is immediately seen from
\mref{det}-\mref{sdet}. These averages were obtained in the
determinant form in \cite{BH}.

\setcounter{equation}{0}
\section{Ladder operators}

We now develop a differentiation formula for the polynomials
$p_n(x)$ orthogonal with respect to the weight $w_0(x)|x-t|^{\ga}$
on the real line, for any smooth reference weight  $w_0$ and  for
general $\ga\geq 0$.
 The derivation given
here is similar to what was previously known \cite{chen1, chen2},
but adapted to the situation where the weight vanishes at one point.

 From the orthogonality condition, there follows the recurrence
relations;
$$
zp_n(z)=p_{n+1}(z)+\al_np_n(z)+\bt_np_{n-1}(z),
$$
with the initial conditions $p_0(z)=1,$ and $\bt_0p_{-1}(z)=0.$ The
diagonal recurrence coefficients can then be expressed as \beq{p1}
\al_n= \textsf{p}_1(n)-\textsf{p}_1(n+1) \eeq where
$\textsf{p}_1(n)$ are defined by expansions \beq{p11}
p_n(z)=z^n+\textsf{p}_1(n)z^{n-1}+... \eeq The coefficients of the
orthogonal polynomials will also have $t$ dependence due to the $t$
dependence of the weight although we denote the polynomials as
$p_n(z)$.

 Since $p_n(z)$ is a polynomial of degree $n$, its derivative
is a polynomial of degree $n-1$ and can therefore be expressed  as a
linear combination of $p_k(z),\;\;k=0,1,...,n-1$,  namely,
\beq{coeff} p_n^{\prime}(z)=\sum_{k=0}^{n-1}C_{n,k}p_k(z). \eeq To
determine the coefficients $C_{n,k}$ we use orthogonality relations
and the formula  \bea
\px|x-t|^{\ga}=\delta(x-t)((x-t)^{\ga}-(t-x)^{\ga})
+\ga\frac{|x-t|^{\ga}}{x-t}. \eea We have
 \bea C_{n,k}&=&\frac{1}{h_k}\Int
p_n^{\pr}(y)p_k(y)w_0(y)|y-t|^{\ga}dy
\nonumber\\
&=&-\frac{1}{h_k}\Int p_n(y)p_k(y)(w^{\prime}_0(y)|y-t|^{\ga}+
w_0(y)\py|y-t|^{\ga})dy\nonumber\\
&=&-\frac{1}{h_k}\Int p_n(y)p_k(y)(\textsf{v}_0^{\pr}(z)-\textsf{v}_0^{\pr}(y))w(y,t)dy
\nonumber\\
&&-\frac{\ga}{h_k}\Int p_n(y)p_k(y)w_0(y)\frac{|y-t|^{\ga}}{y-t}dy
\nonumber\\
&=&-\frac{1}{h_k}\Int
p_n(y)p_k(y)(\textsf{v}_0^{\pr}(z)-\textsf{v}_0^{\pr}(y))w(y,t)dy\nonumber\\
&&-\frac{\ga}{h_k}\Int\frac{p_n(y)p_k(y)}{y-t}w(y,t)dy, \eea where
we used notation $\textsf{v}_0(z)=-\log w_0(z)$.

We note that analogous consideration of $C_{n,n}=0$ implies \beq{f1}
\Int p_n^2(y)\textsf{v}_0^{\pr}(y)w(y,t)dy
=\ga\Int\frac{p_n^2(y)}{y-t}w(y,t)dy, \eeq also the property
$C_{n,n-1}=n$ implies the following Freud equation \beq{34}
n=\frac{1}{h_{n-1}}\Int
p_n(y)p_{n-1}(y)\textsf{v}_0^{\pr}(y)w(y,t)dy
-\frac{\ga}{h_{n-1}}\Int \frac{p_{n}(y)p_{n-1}(y)}{y-t}w(y,t)dy.
\eeq

 Substitution of $C_{n,k}$ into \mref{coeff} and  summation over $k$ using the
Christoffel-Darboux formula;
$$
\sum_{j=0}^{n-1}\frac{p_j(x)p_j(y)}{h_j}
=\frac{p_n(x)p_{n-1}(y)-p_n(y)p_{n-1}(x)}{h_{n-1}(x-y)},
$$
produces the differentiation formula;
\beq{lower}
p_n^{\pr}(z)=-B_n(z)p_n(z)+\bt_nA_n(z)p_{n-1}(z),
\eeq
where
\bea
A_n(z)&:=&\frac{1}{h_n}\Int\frac{\textsf{v}_0^{\pr}(z)-\textsf{v}_0^{\pr}(y)}{z-y}
p_n^2(y)w(y,t)dy+a_n(z,t)\nonumber\\
a_n(z,t)&:=&\frac{\ga}{h_n}\Int \frac{p_n^2(y)}{(y-t)(z-y)}
w(y,t)dy\nonumber\\
B_n(z)&:=&\frac{1}{h_{n-1}}\Int\frac{\textsf{v}_0^{\pr}(z)-\textsf{v}_0^{\pr}(y)}{z-y}
p_n(y)p_{n-1}(y)w(y,t)dy+b_n(z,t)\nonumber\\
b_n(z,t)&:=&\frac{\ga}{h_{n-1}}\Int
\frac{p_n(y)p_{n-1}(y)}{(y-t)(z-y)} w(y,t)dy \eea Equation
\mref{lower} is the ``lowering" operator.

A direct calculation produces two fundamental compatibility conditions valid
for all $z;$
$$
B_{n+1}(z)+B_n(z)=(z-\al_n)A_n(z)-\textsf{v}_0^{\pr}(z)\eqno(S_1)
$$
$$
1+(z-\al_n)(B_{n+1}(z)-B_n(z))=\bt_{n+1}A_{n+1}(z)-\bt_nA_{n-1}(z).
\eqno(S_2)
$$
where we have used \mref{f1} to arrive at $(S_1).$ Without going
into details, we mention here that if the factor $|x-t|^{\ga}$ in
the weight $w(x,t)$ is replaced by
$$
\prod_{j=1}^{\mathcal{N}}|x-t_j|^{\ga_j}
$$
then $(S_1)$ and $(S_2)$ still hold and the only changes are
\bea
a_n(z,t_1,..,t_{\mathcal{N}})&=&
\sum_{j=1}^{\mathcal{N}}\frac{\ga_j}{h_n}
\Int\frac{p^2_n(y)}{(y-t_j)(z-y)}w(y,t_1,..,t_{\mathcal{N}})dy\\
b_n(z,t_1,..,t_{\mathcal{N}})&=&
\sum_{j=1}^{\mathcal{N}}\frac{\ga_j}{h_{n-1}}
\Int\frac{p_n(y)p_{n-1}(y)}{(y-t_j)(z-y)}w(y,t_1,..,t_{\mathcal{N}})dy.\\
\eea Using $(S_1)$ and recurrence relations we have the ``raising"
operator, \beq{raise}
p_{n-1}^{\pr}(z)=(B_n(z)+\textsf{v}_0^{\pr}(z))p_{n-1}(z)-A_{n-1}(z)p_n(z).
\eeq In the next section we take $w_0(x)=\exp(-x^2),$ and make use
of $(S_1)$ and $(S_2)$ to produce a pair of non-linear difference
equations satisfied by the recurrence coefficients for fixed $t.$
These when combined with the $t-$ evolution equations satisfied by
the recurrence coefficients result in a particular Painlev\'{e} IV.

\setcounter{equation}{0}
\section{Derivation of the Painlev\'{e} equation}

For $w(x,t)=\exp(-x^2)|x-t|^{\ga},$ $\textsf{v}_0(x)=x^2$, we find,
\bea
A_n(z)&=&2+a_n(z,t)\\
B_n(z)&=&b_n(z,t). \eea For $z$ near $\infty,$ with fixed $t,$ we
obtain the following asymptotic expansions; \bea
a_n(z,t)&\sim&\frac{2\al_n}{z}+\frac{\ga+2t\al_n}{z^2}+\frac{\ga t+\ga\al_n+2t^2\al_n}{z^3}+..\\
b_n(z,t)&\sim&\frac{2\bt_n-n}{z}+\frac{t(2\bt_n-n)}{z^2}+\frac{\ga\bt_n+t^2(2\bt_n-n)}{z^3}+..,
\eea where the coefficients are determined from orthogonality, the
recurrence relations, \mref{f1} and \mref{34}.

 Substituting the asymptotic expansions into $(S_1)$ and $(S_2),$ we find, by comparing
the coefficients of $1/z^j,$ two difference equations satisfied by
$\al_n$ and $\bt_n;$ \beq{S1}
\bt_{n+1}+\bt_n=n+\frac{1}{2}+\frac{\ga}{2}+\al_n(t-\al_n) \eeq
\beq{S2}
(t-\al_n)\left(\bt_{n+1}-\bt_n-\frac{1}{2}\right)=\bt_{n+1}\al_{n+1}-\bt_n\al_{n-1}.
\eeq

 {\bf Remark \noindent 1.} If $\ga=0,$ then $\al_n=0,$ thus
\mref{S1} and \mref{S2} become $\bt_{n+1}+\bt_n=n+1/2$ and
$\bt_{n+1}-\bt_n=1/2$ respectively. The solution of these equations,
subject to the initial condition $\bt_0=0$ is $\bt_n=n/2,$ which is
the recurrence coefficients of the Hermite polynomials.

{\bf Remark \noindent 2.} If $t=0$, then $\al_n=0,$ then (4.5)
becomes $\bt_{n+1}+\bt_n=n+(1+\ga)/2.$ The unique solution subject
to the initial condition $\bt_0=0$ is $\bt_n=n/2 +\ga(1-(-1)^n)/4,$
which is the recurrence coefficient of what Szeg\"{o} called the
generalized Hermite polynomials (see \cite{Sze}, problem 25). These
should not be confused with the generalized Hermite polynomials
which arise in the rational solutions of Painlev\'{e} IV (see next
section).

To study the $t-$evolution of the recurrence coefficients we begin
by taking a derivative with respect to $t$ of the squared norm $h_n$
of the $n$-th orthogonal polynomial,
 \beq{dh}
\frac{\pt
h_n}{h_n}=\frac{\ga}{h_n}\Int\frac{p_n^2(y)}{t-y}w(y,t)dy=-2\al_n,
\eeq where the last equality is obtained by using relation \mref{f1}
and noting that $\textsf{v}^{\prime}_0(y)=2y.$ Since
$\bt_n=h_n/h_{n-1},$ equation \mref{dh} implies \beq{t1}
\frac{\pt\bt_n}{\bt_n}=2(\al_{n-1}-\al_n). \eeq Differentiating
relation
$$
0=\Int p_{n}(y)p_{n-1}(y)w(y,t)dy,
$$
with respect to $t,$ we find, \bea 0=h_{n-1}\pt \textsf{p}_1(n)+\Int
p_{n}(y)p_{n-1}(y)w_0(y)\pt|y-t|^{\ga}dy,\nonumber \eea where
function $ \textsf{p}_1(n)$ was defined in \mref{p11}. Using the
Freud equation \mref{34} we now get \bea
\pt \textsf{p}_1(n)&=&\frac{\ga}{h_{n-1}}\Int\frac{p_n(y)p_{n-1}(y)}{y-t}w(y,t)dy\nonumber\\
&=&\frac{2}{h_{n-1}}\Int yp_n(y)p_{n-1}(y)w(y,t)dy-n=2\bt_n-n. \eea
In view of relation \mref{p1}, \beq{t2}
\pt\al_n=2(\bt_n-\bt_{n+1})+1. \eeq The equations \mref{t1} and
\mref{t2} are the Toda evolution equations.

We now show that $\tilde{D}_n:=D_n\exp(nt^2)$ satisfies the Toda
molecule equation (c.f. \cite{O}). First note that $$
\sum_{j=0}^{n-1}\al_j=-\textsf{p}_1(n) \quad \mbox{and} \quad
\bt_n=\frac{D_{n+1}D_{n-1}}{D_n^2}. $$ The equation (4.7) together
with (4.9) implies
\bea \pt^2\log
D_n=4\frac{D_{n+1}D_{n-1}}{D_n^2}-2n,\nonumber
\eea and hence
\bea
\pt^2\log
\tilde{D}_n=4\frac{\tilde{D}_{n+1}\tilde{D}_{n-1}}{\tilde{D}_n^2},
\nonumber \eea
which is the Toda molecule equation.

To proceed further, we parameterize $\bt_n$ as \beq{in}
\bt_n=\frac{n}{2}+\frac{r_n}{2}+\frac{\gamma}{4},\;\;\;\;\;r_0=-\frac{\ga}{2},
\eeq then relation \mref{S1} becomes \beq{eq1}
\frac{r_{n+1}+r_n}{2}=(t-\al_n)\al_n. \eeq Multiplying relation
\mref{S2} by $\al_n$ and using the previous relation we get
$$
\frac{r_{n+1}^2-r_n^2}4=\al_n\al_{n+1}\bt_{n+1}-\al_n\al_{n-1}\bt_n.
$$
Therefore
$$
\frac{r_n^2}4=\al_n
\al_{n-1}\left(\frac{n}2+\frac{r_n}2+\frac{\gamma}4\right) +
\textsf{a},$$ where $\textsf{a}$ does not depend on $n$.  Taking
into account the initial condition $ r_0=-\frac{\gamma}{2}$ we
obtain the equation \beq{eq2}
r_n^2=2\left(n+r_n+\frac{\gamma}2\right)\al_n \al_{n-1} +
\frac{\gamma^2}4. \eeq
In terms of the variables $r_n$, the Toda
equations \mref{t1}, \mref{t2} become \beq{eq3}
\al_{n-1}=\al_n+\frac1{2(n+r_n+\frac{\gamma}2)} \pt r_n, \eeq and
 \beq{t3} \pt\al_n=r_n-r_{n+1} \eeq
 respectively. Eliminating $r_{n+1}$
from equations \mref{t3} and (4.12) produces \beq{eq4}
r_n=\al_n(t-\al_n)+\frac12{\pt \al_n}. \eeq To get the differential
equation on $\al_n$ we substitute expressions \mref{eq3}, \mref{eq4}
into (4.13):
$$
\left(\al_n(t-\al_n)+\frac{\al_n'}{2}\right)^2 =
(2n+2\al_n(t-\al_n)+\al_n'+\gamma)\al_n^2 +
\al_n\left(\al_n(t-\al_n)+\frac{\al_n'}{2}\right)'+\frac{\gamma^2}4.
$$
After simplification we obtain the following result.

\begin{theorem}
The recurrent coefficients $\al_n(t)$ satisfy
\beq{pain}
\al_n'' =\frac{\al_n'^2}{2\al_n}+6 \al_n^3 - 8 t \al_n^2 +
2(t^2-\ga-2n-1)\al_n-\frac{\ga^2}{2\al_n}
\eeq
which is a particular fourth Painlev\'{e} equation.
\end{theorem}

\setcounter{equation}{0}
\section{Explicit solutions for even multiplicity}


Painlev\'e IV equation was first represented as a simple system of
three first order equations (dressing chain) in \cite{Sha}. Such a
symmetric form of PIV was used in \cite{NY} to obtain all the
rational solutions of the equation in the remarkable determinant
form (simultaneously with the independent work \cite{KO}). We use
the notations from Noumi-Yamada \cite{NY} to recall their results
and then to use them.

Firstly we bring equation \mref{pain} to the canonical form by a
simple change of variable.
Let $y=2\al_n$ and $\tilde t = -t$ then \mref{pain} takes the form
\beq{p4} y^{\prime\prime}(\tilde t) =\frac{y'^2}{2 y}+\frac32 y^3 +
4 \tilde t y^2 + 2(\tilde t^2-a)y +\frac{b}{y}, \eeq where $a = 2
n+1 +\ga$, $b=-2\ga^2$. Then the symmetric form of PIV  is a system
of first order differential equations satisfied by $f_0=f_0(x),
f_1=f_1(x), f_2=f_2(x)$, where \beq{f} f_1 (x) = - c y (- cx), \eeq
with $c = \sqrt{-3/2}$. The system reads as follows: \bea
f_0^{\pr}+f_0(f_1-f_2)&=&b_0\\
f_1^{\pr}+f_1(f_2-f_0)&=&b_1\\
f_2^{\pr}+f_2(f_0-f_1)&=&b_2
\eea
where
\bea
f_0+f_1+f_2&=&3x\\
b_0+b_1+b_2&=&3, \eea and parameters of the PIV are suitably
expressed in terms of $b_0,b_1$ and $b_2.$ The PIV equation can also
be written in the bilinear form on the level of $\tau$-functions.
The solution of \mref{pain} may then be expressed in terms of
$\tau$-functions $\tau_0(x), \tau_1(x), \tau_2(x)$ as \beq{tau}
f_1=\frac{d}{dx}\log\frac{\tau_2}{\tau_0}+x, \eeq where the
functions $\tau_0$ and $\tau_2$ will be defined later.

 The generalised Hermite
polynomials \cite{NY} are defined as  \beq{schur}
H_{m,n}(x)=\det\left( P_{n-i+j}(x)\right)_{i,j=1}^{m} \eeq where
\bea P_s(x)= \sum_{i+2j=s} \frac{1}{6^j i! j!} x^i. \eea They
coincide with the specialization  $S_{n^m}(x, \frac16,0,0,\ldots)$
of Schur polynomials corresponding to rectangular Young diagrams
containing $m$ rows of length $n$.

 Define also the set of functions
$$
u_{m,n}(x)= \exp\left( -\frac{x^4}{12} + \frac{m-n}2 x^2\right)
H_{m,n}(x).
$$
Then the triple
$$
(\tau_0, \tau_1,\tau_2)=(u_{m,n}, u_{m+1,n}, u_{m,n+1})
$$
leads to a solution of PIV through formulas \mref{p4}--\mref{tau} in the
case $\ga=m$.

\begin{theorem}
The recurrent coefficients $\al_n$ for the weight $w(x)=e^{-x^2}
(x-t)^{2K}$ with $K \in \Z_+$ are given by
$$
\al_n(t)= -\frac12 \frac{d}{d t} \log
\frac{H_{2K,n+1}(t/c)}{H_{2K,n}(t/c)}
$$
where $H_{m,n}(x)$ are defined by \mref{schur}, and $c =
\sqrt{-3/2}$.
\end{theorem}
{\bf Proof.} For $\gamma=2 K$ with $K\in\mathbb{N}$ the orthogonal
polynomials with the weight $w(x,t)$ can be expressed in terms of
Hermite polynomials by the Christoffel formula (\cite{Sze}, pg. 30),
since $w(x,t)$ is the Hermite weight multiplied by a polynomials in
$x.$ It follows from the formula that the recurrence coefficients
$\al_n, \bt_n$ are rational functions of $t$. Therefore $\al_n(t)$
is a rational solution of equation \mref{p4} in this case. The
rational solution of the PIV equation is unique if it exists (see
\cite{GL}) and is expressed in terms of the generalized Hermite
polynomials:
$$
\al_n(t)=\frac12 y(-t) = -\frac{1}{2c} f_1\left(\frac{t}{c}\right)
=-\frac12
\frac{d}{d t} \log \frac{H_{2K,n+1}(t/c)}{H_{2K,n}(t/c)}.
$$
{\bf Remark.} \noindent There is another way to see rationality of
$\alpha_n(t)$. Indeed, equation \mref{sdet} defines an even
polynomials of degree $2Kn$ in $t,$ hence,
$h_{n}(t)=D_{n+1}(t)/D_n(t),$ is rational in $t$ and (4.7) shows
that $\al_n(t)$ is also rational in $t.$

The above considerations allow us to obtain an expression for the
Hankel determinant. We have seen that \mref{dh},
$$
\al_i =-\frac{\pt h_i}{2h_i}.
$$
Therefore
$$
\log h_i -  \log \frac{H_{2K,i+1}(t/c)}{H_{2K,i}(t/c)} = \textrm{const}.
$$
So \beq{st} h_i = a_i \frac{H_{2K,i+1}(t/c)}{H_{2K,i}(t/c)} \eeq for
some constant $a_i$.
\begin{prop} (c.f. \cite{BH}, \cite{FW})
The Hankel determinant for the weight $w(x)=e^{-x^2} (x-t)^{2K}$
with $K \in \Z_+$ is given by
$$
D_n = A_{K,n} H_{2K,n}(t/c)
$$
where $c=\sqrt{-3/2}$, and
$$
A_{K,n}=\prod_{i=0}^{n-1}a_i=
(-1)^{Kn}\pi^{\frac{n}{2}}\frac{3^{Kn}G(2K+n+1)}{2^{Kn+\frac{n(n-1)}{2}}G(2K+1)}
$$
with the Barnes $G-$function \cite{bar} defined by
$$G(z+1)=\Gamma(z)G(z),\;\;\;G(1)=1.$$
\end{prop}
{\bf Proof.} It is clear from \mref{st} and the product expression
of $D_n$,
$$
D_n=h_0 h_1 \ldots h_{n-1}
$$
where $h_i$ are the square of the $L^2$ norm of the monic orthogonal
polynomials, that the constant $A_{K,n}$  in the proposition depends
only on $n, K$, so all we need to do is to determine its value.

 Note that the coefficient of $t^{2Kn}$ of
$D_n(t)$ is equal to the Hankel determinant associated with the
Hermite weight. Therefore \beq{eq11}
 D_n(t) = t^{2Kn}
\pi^{\frac{n}{2}}\prod_{i=0}^{n-1}\frac{i!}{2^i} + \mbox{ lower
order terms}.
\eeq
 On the other hand the leading coefficient of
$H_{2K,n}(t)$ is equal to \beq{eq12} \frac{G(2K+1)G(n+1)}{G(2K+n+1)}
\eeq (see  \cite{NY}). Combining \mref{eq11} and \mref{eq12}
together we get the value of $A_{K,n}$ as stated.

{\bf Remark.} The Hankel determinant $D_n$ as the average of
characteristic polynomial \mref{sdet} was first computed by Brezin
and Hikami in \cite{BH} as determinant of Hermite polynomials. The
equivalence of the resulting formulas with the formulas for the
$\tau$-functions $H_{2K,n}$ for PIV from \cite{NY} was used by
Forrester and Witte in \cite{FW} (see also \cite{KO}). We have now
an explanation for this coincidence through showing that the
diagonal recurrence coefficients $\al_n(t)$ is a solution of  PIV.
We also note that this result can be obtained other way round using
\mref{dh} and  \cite{FW}. 

\vspace{5mm}

 {\bf Acknowledgements}

M.F. is grateful to A.Borodin and A.P.Veselov
for useful discussions. We would like to acknowledge the support of
European research programme ENIGMA (contract MRTN-CT-2004-5652).
M.F. also acknowledges the support of Chapman Fellowship at the
Mathematics Department of Imperial College.

\end{document}